# Luminescence of non-bridging oxygen hole centers as a marker of particle irradiation of α-quartz

L. Skuja[a,*], N. Ollier[b], K. Kajihara[c]

[a]*Institute of Solid State Physics, University of Latvia, 8 Kengaraga str., LV1063, Riga, Latvia*

[b]*Laboratoire des Solides Irradiés CEA-CNRS Ecole Polytechnique, 91128 Palaiseau, France*

[c]*Department of Applied Chemistry for Environment, Graduate School of Urban Environmental Sciences, Tokyo Metropolitan University, 1-1 Minami-Osawa, Hachioji, Tokyo 192-0397, Japan*

## Highlights

- The red emission bands in quartz and silica may be both due to O dangling bonds
- High sensitivity time-resolved detection is possible by pulsed excitation at 266nm
- Damage beyond simple O vacancy formation is needed to form them in quartz
- Their presence in quartz may serve as an indicator for nuclear particle irradiation

## Abstract

The origin of the "red" emission bands in the 600 nm-700 nm region, observed in quartz crystals used for luminescence dating and environmental dosimetry, is still controversial. Their reported spectral and lifetime characteristics are often similar to those of oxygen dangling bonds ("non-bridging oxygen hole centers, NBOHCs") in glassy silicon dioxide. The presence of these "surface radical type" centers in quartz crystal requires sites with highly disordered local structure forming nano-voids characteristic to the structure of glassy $SiO_2$. Such sites are introduced in the tracks of nuclear particles (α-irradiation, neutrons, ions). In case of electrons they are created only at large doses (>5 GGy), approaching amorphization threshold. This study reports a comparison of NBOHC photoluminescence in synthetic quartz and silica glass irradiated by neutrons or 2.5 MeV electrons, and suggests that the red NBOHC photoluminescence band in quartz may serve as a selective marker of an exposure to particle irradiation. It can be distinguished from other red-region luminescence bands by lifetimes in 5-25 µs range, characteristic vibrational structures in the low-temperature spectra and presence of resonance excitation band at ~620 nm.

## Introduction

Quartz is used as a natural luminescent dosimeter in various applications like geological and archeological dating, nuclear accident dosimetry or uranium ore deposits exploration. Luminescence signal readout occurs by thermally (TSL) or, most often, by optically stimulated luminescence (OSL). The dosimetric properties depend not only on the suitable traps for charge carriers, released in TSL or OSL processes but as well on the efficient emitter (recombination) centers. The nature of the emitting centers is still not well-understood.

Different (>10) emission bands observed in quartz and other crystalline $SiO_2$ polymorphs have been reviewed in the context of luminescence dating by Krbetschek et al., 1997; cathodoluminescence (CL) of quartz is reviewed by Götze et al., 2001. The emission band, most often appearing in OSL spectra of quartz is located in ultraviolet (UV) at 3.25 eV (381 nm) (Wintle and Adamiec, 2017). A large amount of evidence relates it to Al substitutional impurity in quartz, however, this assignment is still remains tentative (Martini et al., 2012 and references therein). It is shown that the UV band consists of several independent components (Martini et al., 2014).

The "red" luminescence band located at 1.9 eV (650 nm) is not always present in OSL or TSL, or radioluminescence (Schmidt et al., 2015) spectra of quartz. However, it is a typical feature of ion-induced luminescence of quartz (G.E. King et al., 2011A; Wintle and Adamiec, 2017), its intensity increases with particle irradiation time. The red band is observed in CL spectra of natural quartz at the grain boundaries and in α-particle irradiated regions (Botis et al., 2005; Götze, 2009). In agate crystals the red CL band is the dominant one and grows with irradiation time (Götze et al., 1999). In high quality synthetic α-quartz single-crystals red luminescence is completely absent and is not created by γ-irradiation (Cannas et al., 2004; Kajihara et al., 2013). However, at high irradiation doses by synchrotron X-rays (>20 GGy (G. E. King et al., 2011B)) or 2.5 MeV electrons (7.6 GGy (Skuja et al., 2019)) red luminescence centers are created.

The purpose of this work is to show that the radiation-induced red photoluminescence (PL) centers in *crystalline* $SiO_2$ are in most cases very similar to the defect centers, identified as oxygen dangling bonds in *amorphous/glassy* $SiO_2$, known as "NBOHC" (non-bridging oxygen hole centers) (Griscom, 1991). NBOHCs have been much studied because they are major radiation-induced UV absorbers (Girard et al., 2019; Skuja et al., 2012) in silica optical fibers and other optical devices. Additionally, they are among the very few intrinsic luminescent defects in glassy $SiO_2$, which show zero-phonon lines (Skuja, 1994). NBOHC can be depicted as a surface-type oxygen radical ≡Si−O•, with the "nonbridging" oxygen atom containing unpaired spin (s=1/2) located on the vertice of $SiO_4$ tetrahedron, which is connected by 3 Si-O bonds to the glass network. The PL of NBOHC (≈650nm) can be excited in 2 excitation bands. This paper demonstrates the identification of these PL centers in irradiated crystalline $SiO_2$, using the more efficient excitation by UV laser light into the strong 4.8 eV (258 nm) absorption band of NBOHC, compared to the excitation into the weak 2.0 eV (620 nm) resonance absorption band used in the previous work (Skuja et al., 2019).

## Materials and Methods

High purity (total impurities < 1 µg/g) synthetic α-quartz samples (Asahi Glass/Tokyo Denpa) were irradiated by 2.5 MeV electrons using SIRIUS Pelletron linear accelerator (LSI) with fluences between $1.2 \times 10^{18}$ and $3.0 \times 10^{19}$ $e^-/cm^2$ (0.3 and 7.6 GGy), electron flux $5.5 \times 10^{13}$ $e^-/(cm^2\ s)$ (dose rate 51 MGy/h) and sample temperature 60 °C. They were compared to quartz crystal irradiated by ionizing irradiation at much lower dose (13 MGy $^{60}Co$ γ-rays), and to α-quartz crystal and synthetic dry $SiO_2$ glass, both irradiated by $1 \times 10^{19}$ fast (E>1 MeV) neutrons/$cm^2$. PL spectra and PL kinetics were measured using 4th harmonics of Nd-YAG laser (266 nm) or ArF excimer laser (193 nm) at temperatures between 295 K and 14 K. The room-temperature time-resolved PL spectra were obtained with ~1.5 mJ 10 ns 266 nm pulses repeated at 10 Hz. The temperature dependences and decay kinetics were measured with 0.12 µJ 2 ns 266 nm pulses at 6 kHz. The spectra were recorded by Andor Shamrock 303 spectrographs with 1200, 600 or 150 l/mm gratings and cooled silicon CCD camera (DU971) for CW spectra or image-intensified camera (DH734) for time-resolved spectra. The attainable spectral resolution was 0.25 nm, corresponding to ~0.7 meV in the red spectral region. Kinetics were measured by multichannel photon counter.

## Results

Figure 1 shows time-resolved room-temperature PL spectra of α-quartz crystals irradiated by different doses of ionizing irradiation (0.013 to 7.6 GGy, spectra C,D,E) compared to neutron irradiated ($1 \times 10^{19}$ n/$cm^2$) α-quartz (B) and synthetic $SiO_2$ glass (A). PL was recorded in a 20 μs gate window, delayed by 200 ns after the start of the 10 ns long laser pulse (λ=266 nm, energy ~1.5 mJ). The delay/gate length parameters were selected to maximize the sensitivity, by suppressing the scattered laser excitation light and integrating the PL with lifetimes in 10-20 μs range. The combination of laser excitation with time-resolved detection provides very high sensitivity. The red PL with μs range lifetime is present in all samples except for the α-quartz sample irradiated by the lowest dose (γ-irradiated).

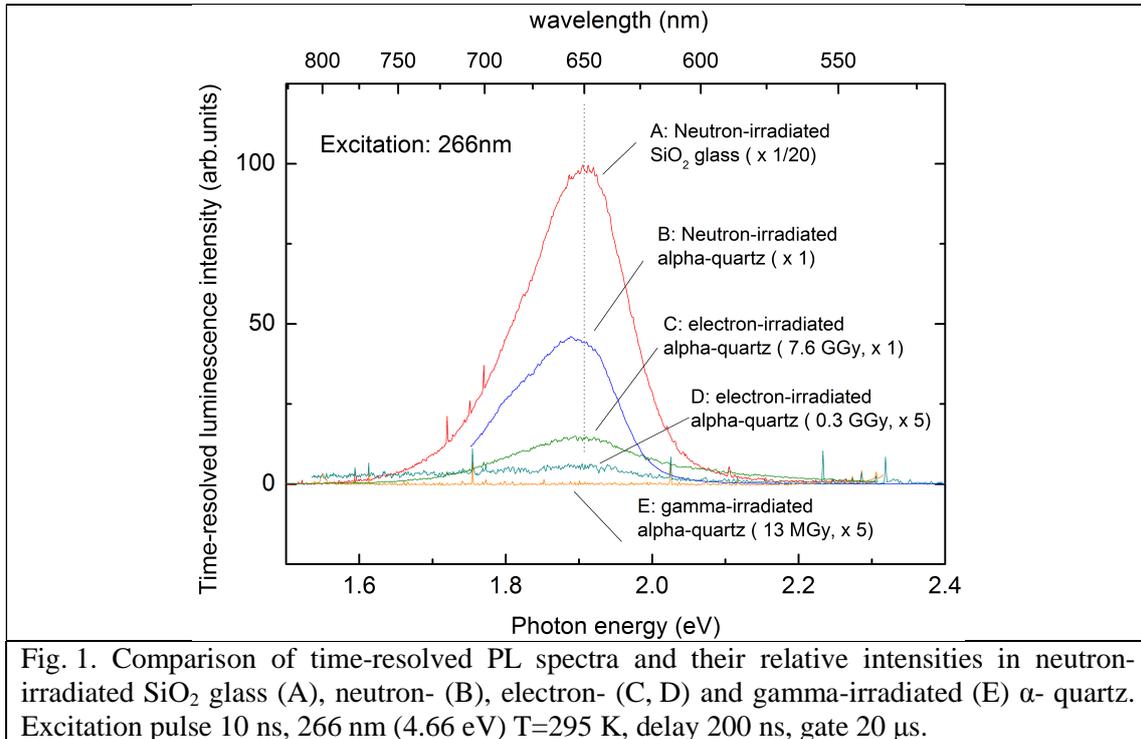

Fig. 1. Comparison of time-resolved PL spectra and their relative intensities in neutron-irradiated $SiO_2$ glass (A), neutron- (B), electron- (C, D) and gamma-irradiated (E) α- quartz. Excitation pulse 10 ns, 266 nm (4.66 eV) T=295 K, delay 200 ns, gate 20 μs.

Figure 2 shows the evolution of the red PL spectra and their intensities in neutron-irradiated (A) and electron-irradiated (B) α-quartz upon cooling from room temperature to 14 K. Insets show the temperature dependence of integrated PL intensities. While the PL spectra in both samples are similar, the temperature dependence is much stronger in electron-irradiated sample (~80 times) compared to the neutron-irradiated quartz (~14 times).

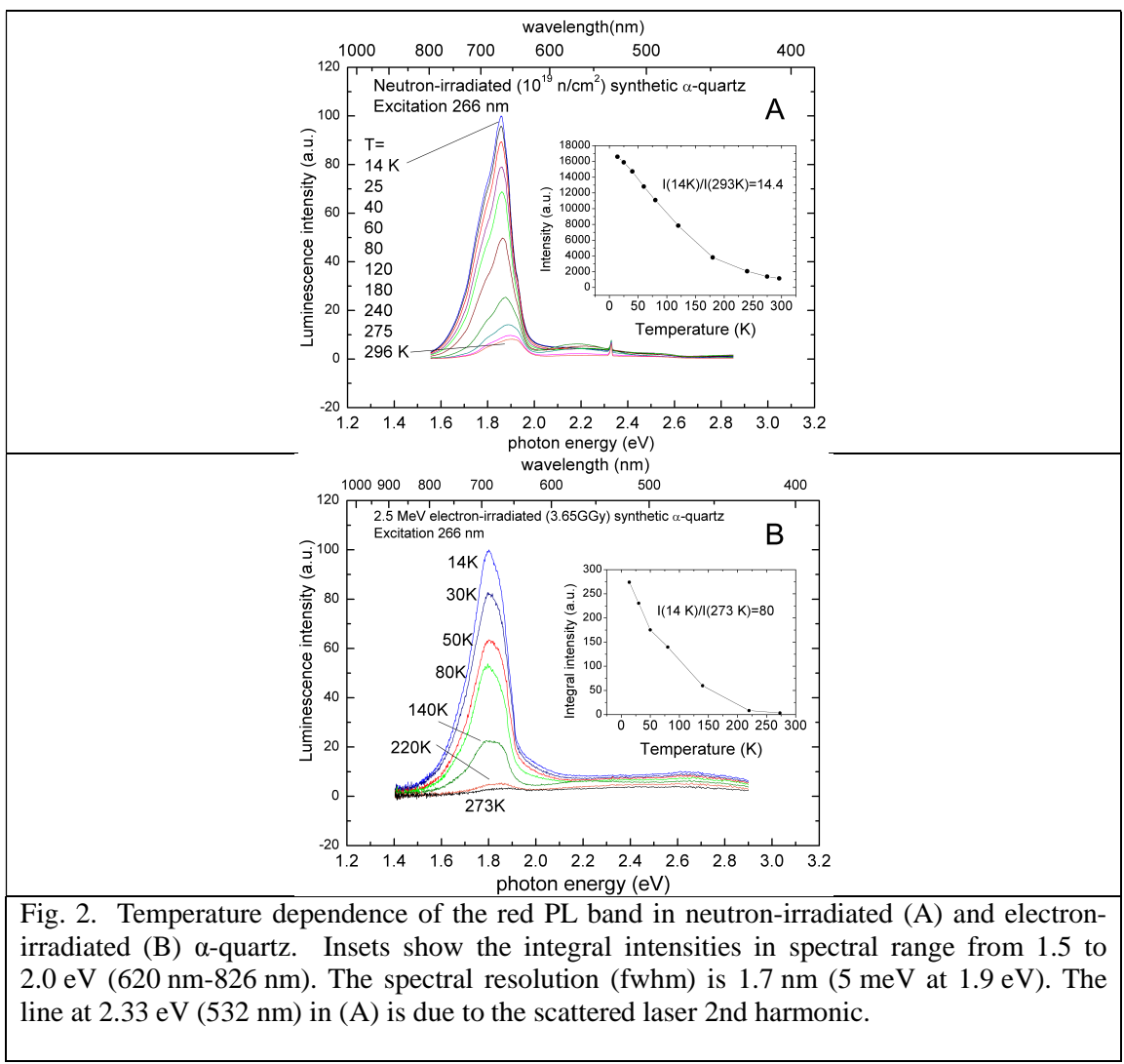

Fig. 2. Temperature dependence of the red PL band in neutron-irradiated (A) and electron-irradiated (B) α-quartz. Insets show the integral intensities in spectral range from 1.5 to 2.0 eV (620 nm-826 nm). The spectral resolution (fwhm) is 1.7 nm (5 meV at 1.9 eV). The line at 2.33 eV (532 nm) in (A) is due to the scattered laser 2nd harmonic.

Figure 3 shows decay kinetics of the red PL and its T-dependence in electron-irradiated quartz, compared to the room-temperature kinetics measured in neutron irradiated quartz or glass. In all cases, both for crystals and glasses, PL decay constant in the 10-20 μs region is observed, both with UV (4.66 eV, 266 nm) or visible (1.96 eV, 633 nm) excitation. In addition, there is a fast component (≤1 μs), observed in all samples, which is the most distinct at room temperature for electron-irradiated crystals (Fig.3 A). It is not depicted for cases of 1.96 eV (633 nm) excitation (Fig.3 B,C) because of the insufficiently fast excitation pulse at this wavelength. However, its presence in neutron-irradiated quartz and silica glass was confirmed in a separate measurement (not shown) using 266 nm 10 ns pulses.

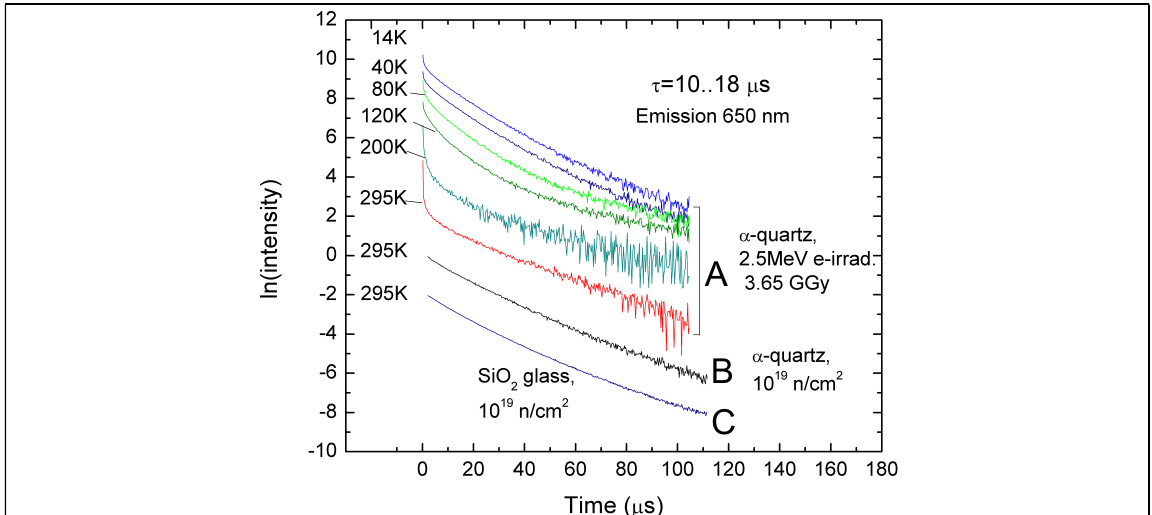

Fig. 3. Decay kinetics of the red PL, measured at 1.91 eV (650 nm) in electron-irradiated α-quartz (A), neutron-irradiated α-quartz (B) and neutron-irradiated SiO$_2$ glass (C). Excitation wavelengths (energies) are 266 nm (4.66 eV) in (A), 633 nm (1.96 eV) in (B), (C). Decay constants in all cases are in 10 μs -18 μs range. Measurement temperatures are indicated.

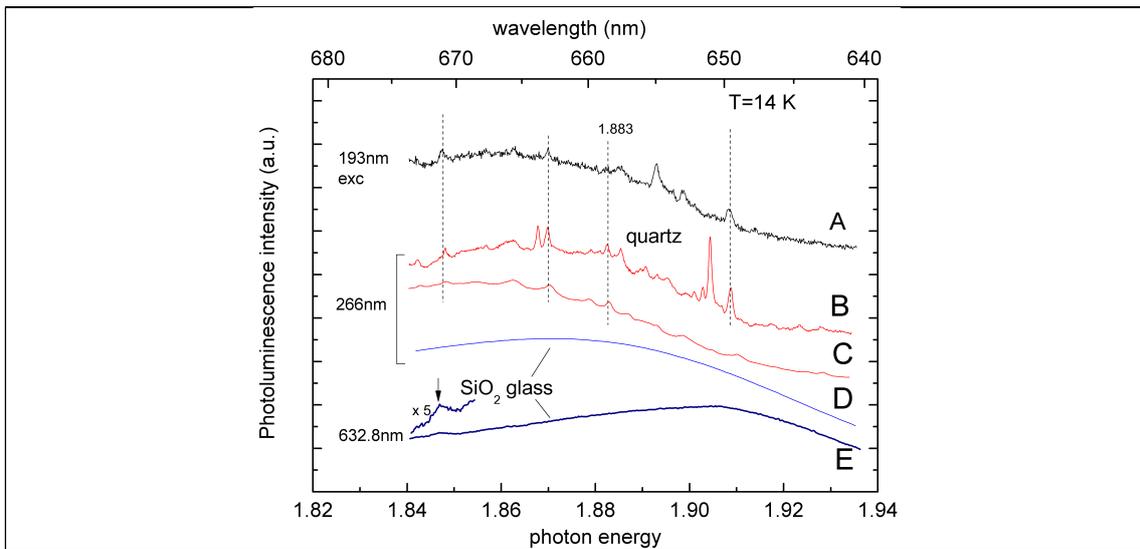

Fig. 4. Comparison of low-temperature (14 K) high-resolution red PL spectra of electron-irradiated α-quartz (3.65 GGy) (A, B), neutron-irradiated (10$^{19}$ n/cm$^2$) α-quartz (C) with neutron-irradiated (10$^{19}$ n/cm$^2$) SiO$_2$ glass (D, E). Excitation wavelengths: 193 nm (A), 266 nm (B, C, D), 632.8 nm (E). Spectral resolution for spectra (A, B, C) 0.24 nm (0.7 meV at 1.9 eV) and 2.5 meV at 1.9 eV for (D) and (E). Spectra are arbitrary normalized and shifted vertically for better viewing. Vertical dashed lines show lines present in several spectra. Arrow shows the vibronic sideband of site-selected zero-phonon line of NBOHC in glass

Figure 4 shows the vibrational structures appearing at low temperature (14 K) in the UV-excited red PL spectra of electron irradiated (A,B) and neutron-irradiated α-quartz (C), and their comparison with PL spectra of neutron-irradiated glass (D,E). The structures are less expressed in neutron-irradiated quartz, compared to electron-irradiated quartz and absent in glass, except for the slight bump at 1.847 eV in spectrum (E), which is the NBOHC Si-(O) vibration sideband (ℏω=0.11 eV) of site-selected zero-phonon line located at the excitation laser photon energy.

## Discussion

The PL spectra of Fig.1 indicate that the red, μs-range PL centers are present in all samples of $SiO_2$, irradiated by particles or by ionizing radiation doses approaching or exceeding the threshold for amorphization of quartz. It starts by ~ 10 GGy and is complete at ~100 GGy (Inui et al., 1990). The absence of this PL in un-irradiated or in γ- irradiated (<1 GGy) quartz crystals is in accord with the previous observations (Cannas et al., 2004; Kajihara et al., 2013). The presence of this PL even in the highest purity quartz crystals indicates its intrinsic character. This makes unlikely the contribution of impurity ions like $Fe^{3+}$, which emit in the red region (1.76 eV (704 nm), (Götze et al., 2001)).

A number of distinctive features:

- the position of emission spectra (~peak 1.9 eV (652 nm), fwhm ~0.17 eV-0.2 eV at 295 K) (Fig.1);
- the decay lifetimes in 10 μs – 20 μs region, which are relatively little dependent on temperature, and do not follow the temperature dependences of intensity (Figs. 2,3);
- the presence of weak nearly resonance PL excitation band at ~ 2 eV (620 nm) and of ~100 times stronger band at 4.8 eV (258 nm) and higher energies (Skuja et al., 2012);
- The presence of sharp vibronic lines in low-temperature PL spectra, observed by site-selective excitation in glass (Skuja et al., 2012) or by non-selective excitation in heavily irradiated quartz crystals (Fig. 4);

indicate that the red PL observed in irradiated quartz is similar to the PL of oxygen dangling bonds (NBOHC's) in glassy $SiO_2$. Due to their impact on optical properties of optical fibers (Girard et al., 2019) and silica UV optics, NBOHCs have been studied in detail by EPR (Griscom, 1991) and optical spectroscopy (Skuja, 1994). The vast evidence accumulated on NBOHCs in glassy $SiO_2$ is reasonably directly applicable as well to quartz.

NBOHC's are surface-type centers. They are detected also as oxygen radicals on surface of $SiO_2$ (Vaccaro et al., 2008). In glassy $SiO_2$, nanosized voids in the network of corner-shared $SiO_4$ tetrahedra exist, as indicated by significantly lower density of glass (2.20 g/cm$^3$) vs. quartz (2.65 g/cm$^3$). The "bulk" NBOHCs can form on these internal surfaces. Such surfaces do not exist in an intact α-quartz lattice. Hence, the detection of NBOHC luminescence in irradiated crystalline $SiO_2$ indicates creation of such surfaces – either by local disorder around nuclear particle tracks or by amorphization due to ultra-high doses (> 1 GGy) of ionizing irradiation. Doses of such magnitude, however, are uncommon for quartz dosimetry tasks. At lower ionizing radiation doses, only oxygen vacancies and interstitials form (~$5\times10^{16}$ vacancies/cm$^3$ after 0.05 GGy, and NBOHC luminescence is absent (Kajihara et al., 2013)). Therefore the presence of the NBOHCs luminescence in quartz can be considered as an indicator of nuclear particle (e.g., α-particle) irradiation. The sensitivity of luminescence technique, when pulsed laser excitation and gated photon counting are used, is commonly very high, NBOHCs created along single nuclear track should be detectable. Quantitative data relating luminescence intensity and irradiation dose parameters are presently not available.

The suggested relationship between particle irradiation and the presence of NBOHCs in quartz assumes an "ideal" crystal quartz lattice as a starting point. However, in real crystals, in particular in natural samples, a complicating factor arises from hydrogen- or alkali-passivated precursors in less-ordered sites, e.g., on grain surfaces, where oxygen dangling bonds ≡Si−O• can be possibly created without nuclear damage: by radiolysis of ≡Si−O−H or ≡Si−O-(alkali ion) species. Such impurity related, extrinsic mechanism of NBOHC generation is well-known in glassy $SiO_2$. This process is probably efficient in high SiOH-content microcrystalline $SiO_2$, like opal or agate, where intense 1.9 eV (652 nm) CL is observed (Götze et al., 1999). CL and electron paramagnetic resonance studies indicate that defect centers are located mostly in grain surface regions (Botis et al., 2005; Timar-Gabor, 2018). It can be suggested that the distinction between the "intrinsic", particle irradiation-generated NBOHCs in quartz and "extrinsic" NBOHCs, created from impurity-related precursors, can be made by thermal annealing and re-irradiation by purely ionizing irradiation (X-rays or γ-rays): only the extrinsic NBOHCs should be re-introduced. An additional evidence could be provided by the amount of inhomogeneous broadening in PL spectra, which is larger in the case of nuclear particle-irradiated quartz (Skuja et al., 2012), (Fig. 4, compare spectra B and C), and should be smaller in centers, generated from extrinsic precursors in ordered crystalline environment. The verification of these assumptions is a subject to further study.

## Conclusion

The luminescence centers in $SiO_2$ crystalline polymorphs, which give rise to the red luminescence band with peak at ~1.9 eV (~650 nm) and lifetime in 10-20 μs range in many cases are the analogs of oxygen dangling bonds ("NBOHCs") in amorphous $SiO_2$. Local nanovoids in α-quartz structure are necessary to form them. This can occur by local amorphization in the tracks of nuclear particles; a simple O vacancy-interstitial mechanism, initiated by pure ionizing irradiation is not sufficient. If luminescent centers on the grain boundaries can be excluded, then the luminescence of NBOHC in quartz can serve as a sensitive indicator of nuclear particle irradiation.

**Acknowledgement**. This work was supported by Latvian Science Council project lzp-2018/1-0289. L.S. was additionally supported by the visiting professor program of Ecole Polytechnique. K.K. was partially supported by the Collaborative Research Project of Laboratory for Materials and Structures, Tokyo Institute of Technology. The input of referees is highly appreciated.